\documentclass[12pt]{article}
\usepackage{mathrsfs}
\usepackage{amsmath,amssymb,array}
\usepackage{graphicx}
\usepackage{bbm}
\usepackage{multicol}
\newcommand{\cN}{{\cal N}}
\numberwithin{equation}{section}
\def\p{\partial}

\begin{document}

\begin{titlepage}
\renewcommand{\thefootnote}{\fnsymbol{footnote}}

\begin{center}
{\LARGE \bf Mesons Mass Spectrum in a Modified Soft-Wall AdS/QCD Model}

\vspace{1.0cm}

{Keita Kaniba Mady$^{\dag,\star}$}\footnote{E-mail: madyfalaye@gmail.com},
{Dicko Younouss Ham\`{e}ye$^{\dag,\ddag}$}
\\

\vspace{.5cm}
{\it\small {$^{\star}$ Centre de Calculs, de Mod\`{e}lisations et de Simulations, FAST, Bamako,\\
$^{\dag}$ D\'{e}partement de Physique, Facult\'{e} des Sciences et Techniques, Universit\'{e} des Sciences, des Techniques et des Technologies de Bamako, Mali,}\\ $^{\ddag}$ Ecole Superieure de Technologie et de Management, ESTM, Mali}\\

\vspace{.3cm}
\today
\end{center}\vspace{1.5cm}

\centerline{\textbf{Abstract}}\vspace{0.5cm}

Undeniably, Quantum Chromodynamics is not a Conformal Field Theory (CFT) at the mesons mass scale. So one would undoubtedly wonder why the Anti-de Sitter Space/ Conformal Field Theory (AdS/CFT) correspondence has been applied to it. Regarding AdS/CFT duality as an indication to a duality much more applicable to QCD, we concoct in this letter a possible gravity dual to QCD by considering its non-conformal feature.  Surprisingly, the agreement between the mass spectrum of $\rho$ vector messon compare to its experimental value is spectacular.

\end{titlepage}
\setcounter{footnote}{0}


\section{Introduction}

The Anti-de Sitter Space/ Conformal Field Theory (AdS/CFT) correspondence is currently one of our best mathematical framework to study strongly conformal coupled gauge theories. Popularized after the work of Maldacina \cite{M}, it is linked  to the old idea of gauge/gravity duality advised by 't Hooft \cite{tH}. The novelty of AdS/CFT duality is that we can, for the first time, predict what we want about a d-dimensional strongly conformal gauge theory by simply investigating a dual gauge theory living on a (d+1)- dimensional AdS spacetime.

Quantum Chromodynamics is a non-Abelian gauge theory, based on the $SU(3)_{c}$  gauge group, that describes the physics of the hadrons(that is the systems of quarks interacting among themselves via the exchange of the $SU(3)_{c}$ gauge fields). The quarks are charged under the action of the $SU(3)_{c}$  gauge group, and their charges are called colors.

The hadrons are divided in two blocks: the baryons (bounded state of three quarks states with net color charge being zero) and the mesons (bounded state of quark-antiquark).
The mesons as well as the baryons are all massive and confined inside the hadrons with a confinement scale $\Lambda_{QCD}$. Therefore, QCD is not a conformal field theory. So the final outcome is that QCD  is neither a conformal field theory, nor a supersymmetric field theory.

 The AdS/CFT duality, as original propounded by Maldacina \cite{M}, preaches the equivalence between the low energy approximation of type $IIB$ string theory on $AdS_{5}\bigotimes S^{5}$ and  $\cN=4$ U(N) Supersymmetry Yang Mills theory for large $N$ in four dimensions. This Anti-de Sitter/conformal field theory (AdS/CFT) correspondence\cite{GKP, OGO, W}, has been widely studied at length in the literature.

At first sight, there is no indications that AdS/CFT duality can be applied to QCD due to the non-conformal and the non-supersymmetric aspects of QCD. However, there are presently two routes in the bottom up level to study QCD through AdS/QCD duality: The hard-wall model \cite{dTB, EKSS, DP1} and the soft-wall model \cite{KKSS}.

In these models, AdS/QCD correspondences has been regarded as a device for learning about the low energy properties of QCD such as the mass spectrum, the form factors, the correlation functions, and the decay constant of the bounded state of quarks.

In this paper, we modified the current soft-wall model by considering the non-conformal property of low energy QCD. Specifically, we investigate the mass spectrum of the vector ($\rho$) mesons in this modified version. The main feature of soft-wall AdS/QCD model is that within it one can easily achieve linear confinement \cite{KKSS, Z, SWY} and chiral symmetry breaking \cite{GKK, DP1, VS2}.
The same calculations can be performed by using our modified version. The fundamental lesson preached by this AdS/QCD duality is still the same , that is, the equivalence between gravity theory living in a five-dimensional AdS geometry and low energy QCD at the boundary of this geometrical background.

Most analysis in the soft-wall model \cite{KKSS, Z, VS1, VS2, GKK, SWY, WF, KD} have been devoted to the mass spectrum of the resonance messons sector of the low energy QCD. Perusal of any of these models reveals that the discrepancy between the theoretical value and the experimental value of some of the masses is higher than $10\%$. In this manuscript reducing those discrepancies is behind our motivation. As can be seen in this paper, we reduce considerably those discrepancies in a very spectacular way.

The rest of this paper is organized as follows: The next section  reviews the soft-wall AdS/QCD, and meantime presents our parameterizations and the mass spectrum. The final section is devoted to the conclusion.

\section{The Soft-Wall AdS/QCD and Its Modified version}

In the soft wall AdS/QCD \cite{GKK, Z, VS1, KKSS, KKSS2} , the geometry in which the bulk fields propagate is assumed to be 5D AdS space with the metric given in the Poincar\'{e} coordinate by
\begin{eqnarray}
ds^2=\,G_{MN}\,dx^M dx^N=a(z)\,(\,\eta_{\mu\nu}dx^{\mu}dx^{\nu}+dz^2),\label{metric},
\end{eqnarray} where $(\eta_{\mu\nu})=diag(1,-1,-1,-1)$ is the 4D Minkowski metric, and $a(z)$ is the conformal factor or the warped factor. The conformal coordinate $z$ within the range $0<z<\infty$ sets the energy scale in the 4D Minkowski space.

The dual theory of QCD in soft wall models contains the $SU(N_f)_L \times SU(N_f)_R$ gauge fields (A$_{L}$ and A$_{R}$) and a scalar field $X(z)$ field, which belong to the adjoint representation of the $SU(N_f)_L \times SU(N_f)_R$ gauge group. The 5D action that seems to describe the $\rho$ vector mesons of QCD from these fields reads
\begin{eqnarray}
S_M=\int d^4x\,dz\, \sqrt{G}\,e^{-\Phi}\, \mathrm{Tr}\left\{-\frac{1}{4g_5^2}(\,F_L^2+F_R^2)+
    |DX|^2-m_X^2|X|^2\,\right\}\, \label{SM},
\end{eqnarray}
where $F_L$ and $F_R$ are the non-abelian field strength and the symbol $D$ is the Yang-Mills covariant derivative all are defined in \cite{KD}. The dilaton field is assumed to be dependent on the holographic coordinate $z$, that is, $\Phi\sim\Phi(z)$. The vacuum expectation value $\langle{X}\rangle=\frac{1}{2}\,v(z)$ of the scalar field $X$ is responsible of the spontaneous breaking of the chiral symmetry of low energy QCD and it satisfies $v(z\rightarrow0)\sim \,\alpha z+\beta z^3\,.$

The soft-wall AdS/QCD \cite{EKSS} from its inception has been built in the philosophy that the combination $\Phi-\log[a]$ satisfies two constraints:
\begin{enumerate}
\item $(\Phi-Log[a])(z\rightarrow\infty)\sim \,\lambda^2 z^2\, $  when one Requires the dual theory to reproduce linear confinement aspect of QCD.
\item $(\Phi-Log[a])(z\rightarrow0)\sim \,\log[z]\, $ which is a consequent of satisfying the conformal symmetry at UV.
\end{enumerate}

In this paper, our strategy is to relax the second constraint, since QCD is simply not a conformal theory at the energy scale under consideration here. Linear confinement is an undeniable property of low energy QCD and has been well established experimentally, so our model will be only based on the first constraint. The mapping between the bulk fields and their dual in QCD is given in table $1$.
\begin{table}
\begin{center}
\caption{The mapping between the fields in the two side and their dimensions.}
\begin{tabular}{ c c c | c c c | c c c | c c c }
  \hline
  \hline
  & 4D : \textit{O}(x) & & & 5D : $\Phi^{bulk}(x,z)$ & & & $\Delta$ & & & $m_{5}^{2}$ & \\
  \hline
  & $\overline{\psi}_{L} \gamma^{\mu} t^{a} \psi_{L}$ & & & $A_{L \mu}^{a}$ & & & 3 & & & 0 & \\
  & $\overline{\psi}_{R} \gamma^{\mu} t^{a} \psi_{R}$ & & & $A_{R \mu}^{a}$ & & & 3 & & & 0 & \\
  & $\overline{\psi}_{R}^{\alpha} \psi_{L}^{\beta}$ & & & $\frac{1}{z} X^{\alpha\beta}$& & & $3$ & & & $-3$ & \\
  \hline
  \hline
\end{tabular}
\end{center}
\end{table}

The hadrons in soft wall AdS/QCD models are identified as the renormalizable modes of the 5D gauge fields. Concretely, the vector mesons (V) which the subject of this manuscript is  defined by the relation:
\begin{eqnarray}
  V=(A_{L}+A_{R})/2\,.
 \nonumber
\end{eqnarray}

The decomposition of this vector field via the Kaluza-Klein mechanism is usually done by $V_\mu(x,z)=\sum_{n}\,\rho_\mu^{(n)}(x)h_V^{(n)}(z)$. The infinite tower of 4D massive vector fields $\rho_\mu^{(n)}(x)$ called the (KK) modes are then identified as the vector mesons of the ordinary low energy QCD. The mass spectrum of this vector mesons is derived from the constraint satisfied by the holographic coordinate dependent parts \cite{KM} which in the axial gauge, $V_z=0$,
 reads
\begin{equation}\label{eigen}
    -\p_5(ae^{-\Phi}\p_5h_V^{(n)})=ae^{-\Phi}M_V^{(n)2}h _V^{(n)},
\end{equation} in which $M_V^{(n)2}$ are the masses of the vector fields $\rho_\mu^{(n)}(x)$.

Equation (\ref{eigen}) can be rewritten into a Schr\"{o}dinger form , by using the transformation $h_V^{(n)}=e^{[\Phi(z)-\log{a(z)}]/2}\chi_V^{(n)}$ and the outcome is
\begin{equation}\label{sch}
  -\chi_V^{(n)\prime\prime}+V_V\chi_V^{(n)}=M_V^{(n)2}\chi_V^{(n)}
\end{equation}
 with the potential given by
\begin{eqnarray}
V_V=\frac{1}{4}[\Phi(z)-\log{a(z)}]'^{\,2}-\frac{1}{2}[\Phi(z)-\log{a(z)}]''\,,\label{VV}
\end{eqnarray} where ($'$) denotes derivation with respect to z.
Now having review the soft wall AdS/QCD model, we propose the following parameterizations for the dilaton field and the  warp factor:
\begin{eqnarray}
  \Phi(z)=\lambda^2 z^2+\lambda z-\log(z)\,,\qquad  a(z)=1/z\,,\label{para}
\end{eqnarray}

To solve equation (\ref{sch}) together with (\ref{para}), we use the shooting method with the boundary conditions $\chi_n(z\to 0) =0$, $\partial_z \chi_n(z\to \infty) =0$. The resulting mass spectrum of the $\rho$ vector mesons with $ \lambda=423\,\mathrm{MeV}\,$ is presented in table $2$.
\begin{table}
\begin{tabular}{|c|c|c|c|c|c|c|c|c|}
\hline
$\rho$  mass (MeV)                  &  0  &  1   &  2   &  3   &  4   &  5   &  6   \\
\hline
Model  &771.8& 1188.6 & 1484.7 & 1727 & 1937.5 & 2126  & 2298 \\
\hline
experimental   &775.5$\pm$ 1& 1282$\pm$ 37 & 1465$\pm$ 25 & 1720$\pm$ 20 & 1909$\pm$30 & 2149$\pm$ 17 & 2265$\pm$ 40 \\
\hline
error  &0.5\%& 7.3\% &1.3\% & 0.4\% & 1.5\% & 1.1\% & 1.5\% \\
\hline
\end{tabular}
\caption{\small{The theoretical and experimental values of the masses of the vector $\rho$ meson.}}\label{rho}
\end{table}

\section{Conclusions}

In this manuscript, we modified the current soft-wall AdS/QCD models by considering the non-conformal aspect of QCD. The predicted mass spectrum of the $\rho$ mesons agrees surprisingly well with its experimental values. It is therefore important to extend this analysis to the other sectors.

\subsection*{Acknowledgments}
Keita is indebted to all the members of the department of physics of Nanchang University, and the teams of the Center for Relativistic Astrophysics and High Energy Physics of Nanchang University for their hospitality.


\newpage
\begin{figure}[t]
\begin{center}
\includegraphics[width=20cm,clip=true,keepaspectratio=true]{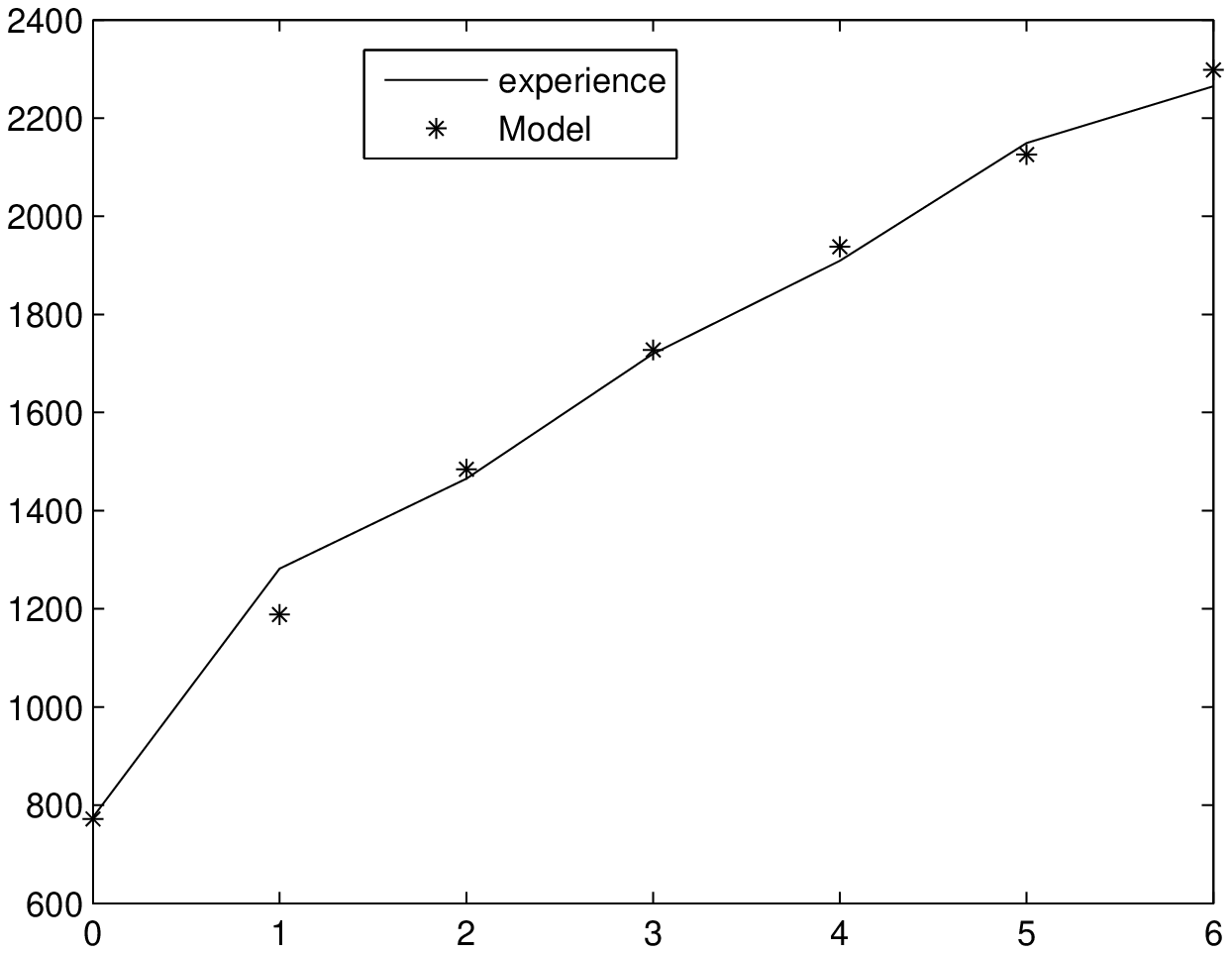}
\caption{\small The mass spectra of the different model compare to the experimental data.}
\end{center}\label{kk}
\end{figure}

\begin{thebibliography}{99}
\bibitem{tH} G. 't Hooft, "A Planar Diagram Theory for Strong Interactions,"
             Nucl. Phys. B72, 461 (1974).
\bibitem{M}  J. M. Maldacena, "The Large N Limit of Superconformal Field Theories and Supergravity,"
             Adv. Theor. Math. Phys. 2, 231 (1998) [arXiv:hep-th/9711200].
\bibitem{GKP} S. S. Gubser, I. R. Klebanov and A. M. Polyakov, "Gauge Theory Correlators from Non-Critical String Theory,"
             Phys. Lett. B428, 105 (1998) [arXiv:hep-th/9802109].
\bibitem{W}  E. Witten, "Anti De Sitter Space And Holography",
             Adv. Theor. Math. Phys. 2, 253 (1998) [arXiv:hep-th/9802150].

\bibitem{dTB} G. F. de T\'{e}ramond and S. J. Brodsky, "Hadronic Spectrum of a Holographic Dual of QCD,"
              Phys. Rev. Lett. 94, 0201601 (2005) [arXiv:hep-th/0501022].
\bibitem{EKSS} J. Erlich, E. Katz, D. T. Son and M. A. Stephanov, "QCD and a Holographic Model of Hadrons,"
              Phys. Rev. Lett. 95, 261602 (2005) [arXiv:hep-ph/0501128].
\bibitem{OGO}  O. Aharony, S.S. Gubser, J. Maldacena, H. Ooguri, Y. Oz, "Large N Field Theories, String Theory and Gravity,"
              Phys. Rept. 323, 183 (2000) [arXiv:hep-th/9905111].
\bibitem{DP1} L. Da Rold and A. Pomarol, "Chiral Symmetry Breaking from Five Dimensional Spaces,"
              Nucl. Phys. B721, 79 (2005) [arXiv:hep-ph/0501218].
\bibitem{KM}Thomas M. Kelley,  "The Dynamics and Thermodynamics of Soft-Wall AdS/QCD," [arXiv:hep-ph/1108.0653].
\bibitem{KKSS} A. Karch, E. Katz, D. T. Son and M. A. Stephanov, "Linear Confinement and AdS/QCD",
              Phys. Rev. D74, 015005 (2006) [arXiv:hep-ph/0602229].
\bibitem{GKK} T. Gherghetta, J. I. Kapusta and T. M. Kelley, "Chiral Symmetry Breaking in Soft-Wall AdS/QCD,"
              Phys. Rev. D79: 076003 (2009) [arXiv:0902.1998].
\bibitem{Z} P. Zhang, "Linear Confinement for Mesons and Nucleons in AdS/QCD",
             JHEP 05 (2010) 039 [arXiv:1003.0558]; "Mesons and Nucleons in Soft-Wall AdS/QCD,"
             Phys. Rev. D82, 094013 (2010) [arXiv:1007.2163]; "Constraining the Infrared Behavior of the Soft-Wall AdS/QCD Model,"
             [arXiv:1105.6293].
\bibitem{VS1} A. Vega and I. Schmidt, "Hadrons in AdS/QCD Correspondence,"
              Phys. Rev. D79, 055003 (2009) [arXiv:0811.4638].
\bibitem{VS2} A. Vega and I. Schmidt, "Modes with variable mass as an alternative in AdS / QCD models with chiral symmetry breaking,"
              Phys. Rev. D82, 115023 (2010) [arXiv:1005.3000].
\bibitem{KD} K.M.KEITA, and Y.H.DICKO, "Note on the Mesons Mass Spectrum in a Soft-Wall AdS/QCD Model"[arXiv:1112.3204].
\bibitem{SWY}  Yan-Qin Sui, Yue-Liang Wu, Zhi-Feng Xie, Yi-Bo Yang , "Prediction for the Mass Spectra of Resonance Mesons in the Soft-Wall AdS/QCD with a Modified 5D Metric,"
              Phys. Rev. D81, 014024 (2010) [arXiv:0909.3887].
\bibitem{WF} Jonathan P. Shock and Feng Wu, "Three Flavour QCD from the Holographic Principle," JHEP 0608:023 (2006) [arXiv:0603142]; Jonathan P. Shock, Feng Wu, Yue-Liang Wu,and Zhi-Feng Xie, "AdS/QCD Phenomenological Models from a Back-Reacted Geometry," JHEP 0703:064 (2007) [arXiv:0611227].
\bibitem{HG} Hongying Jin and Gang Liu, "An improved dilaton Background in soft wall model,"
                JHEP 1011:147 (2010) [arXiv:1009.3548].

\bibitem{KKSS2} A. Karch, E. Katz, D. T. Son and M. A. Stephanov, "On the sign of the dilaton in the soft wall models,"
                JHEP 04 (2011) 066 [arXiv:1012.4813].

\end{thebibliography}
\end{document}